\documentclass[aps,prx,twocolumn,groupedaddress]{revtex4-2}
\usepackage{amsmath}
\usepackage{amssymb} 
\usepackage[group-separator={,}]{siunitx}
\usepackage{mathtools} 
\usepackage{hyperref} 
\usepackage{xcolor}
\usepackage{quantikz}
\usepackage{algorithm}
\usepackage{algpseudocode} 
\usepackage[T1]{fontenc}
\usepackage[normalem]{ulem}
\usepackage{graphicx}
 
\begin{document} 
\title{Scalable Simulation of Quantum Many-Body Dynamics with Or-Represented Quantum Algebra}
\author{Lukas Broers$^{1}$} 
\email{Lukas.broers@riken.jp} 
\author{Rong-Yang Sun$^{1,2}$} 
\author{Seiji Yunoki$^{1,3,4,5}$}
\affiliation{$^{1}$Computational Materials Science Research Team, RIKEN Center for Computational Science (R-CCS), Kobe, Hyogo, Japan} 
\affiliation{$^{2}$RIKEN Interdisciplinary Theoretical and Mathematical Sciences Program (iTHEMS), Wako, Saitama, Japan} 
\affiliation{$^{3}$Quantum Computational Science Research Team, RIKEN Center for Quantum Computing (RQC), Wako, Saitama, Japan} 
\affiliation{$^{4}$Computational Condensed Matter Physics Laboratory, RIKEN Cluster for Pioneering Research (CPR), Wako, Saitama, Japan} 
\affiliation{$^{5}$Computational Quantum Matter Research Team, RIKEN Center for Emergent Matter Science (CEMS), Wako, Saitama, Japan} 
\date{\today}
\begin{abstract} 
High-performance numerical methods are essential not only for advancing quantum many-body physics but also for enabling integration with emerging quantum computing platforms. 
We present a scalable and general-purpose parallel algorithm for quantum simulations based on or-represented quantum algebra (ORQA).
This framework applies to arbitrary spin systems and naturally integrates with quantum circuit simulation in the Heisenberg picture, 
particularly relevant to recent large-scale experiments on superconducting qubit processors [Kim {\it et al.}, Nature {\bf 618}, 500 (2023)]. 
As a benchmark, we simulate the kicked Ising model on a 127-qubit heavy-hexagon lattice, tracking the time evolution of local magnetization using up to one trillion Pauli strings. 
Executed on the supercomputer Fugaku, our simulations exhibit strong scaling up to $2^{17}$ parallel processes with near-linear communication overhead. 
These results establish ORQA as a practical and high-performance tool for quantum many-body dynamics, and highlight its potential for integration into hybrid quantum-classical computational frameworks, complementing recent advances in tensor-network and surrogate simulation techniques. 
\end{abstract} 

\maketitle

\section{Introduction}

Despite continuous progress in the development of quantum hardware, the milestone of fault-tolerant quantum computing (FTQC)--and with it, definitive quantum supremacy--has yet to be achieved~\cite{arute2019quantum,liu2021closing,kim23utility,googleError}.
With the steadily improving capabilities of noisy intermediate-scale quantum devices~\cite{Preskill2018,RevModPhysNisq}, the hybrid integration of quantum processors with high-performance computing (HPC) pipelines has emerged as a promising strategy for realizing quantum advantage before the FTQC era. 
Recent advances in quantum-centric supercomputing architectures exemplify this direction~\cite{Alexeev2024,javier24}. 
At the same time, pre-fault-tolerant quantum processors have demonstrated practical utility through accurate expectation value measurements, enabled by sophisticated error mitigation techniques~\cite{kim23utility}. Remarkably, these results have been reproduced--and in some cases even surpassed--by state-of-the-art classical simulation methods~\cite{liao23,rudolph2023,Tindall24,Orus24,begusic23, begusic24}.
In parallel, developments in quantum machine learning have highlighted the potential of hybrid quantum-classical optimization loops, suggesting further opportunities for synergy between quantum hardware and classical HPC infrastructures~\cite{QML17,Cerezo22}.

This class of hybrid quantum approaches relies equally on progress in classical high-performance computation and on the development of more capable quantum processors.
Among the many computational techniques developed to address quantum many-body systems, tensor network methods stand out as particularly powerful.  
In particular, matrix product states (MPS)~\cite{white1992density,ostlund1995thermodynamic,DMRG11,paeckel2019time} and projected entangled pair states (PEPS)~\cite{verstraete2004renormalization,verstraete2004valence} have become foundational tools in computational quantum physics~\cite{Orus19, Cirac21}. 
While these methods were not originally designed to interface with quantum hardware, they have proven indispensable for benchmarking, verifying, and interpreting results from quantum simulation experiments. In addition, recent developments in surrogate simulation techniques, such as sparse Pauli dynamics and operator-based classical simulation frameworks~\cite{rudolph2023,fontana2023,begusic24b}, offer complementary strategies that scale favorably with certain circuit structures or observables. These approaches are helping to bridge the gap between classical and quantum simulations, providing practical tools for interpreting experimental results and guiding algorithmic design. 

\begin{figure*}[h!t]  
    \includegraphics[width=1.0\linewidth]{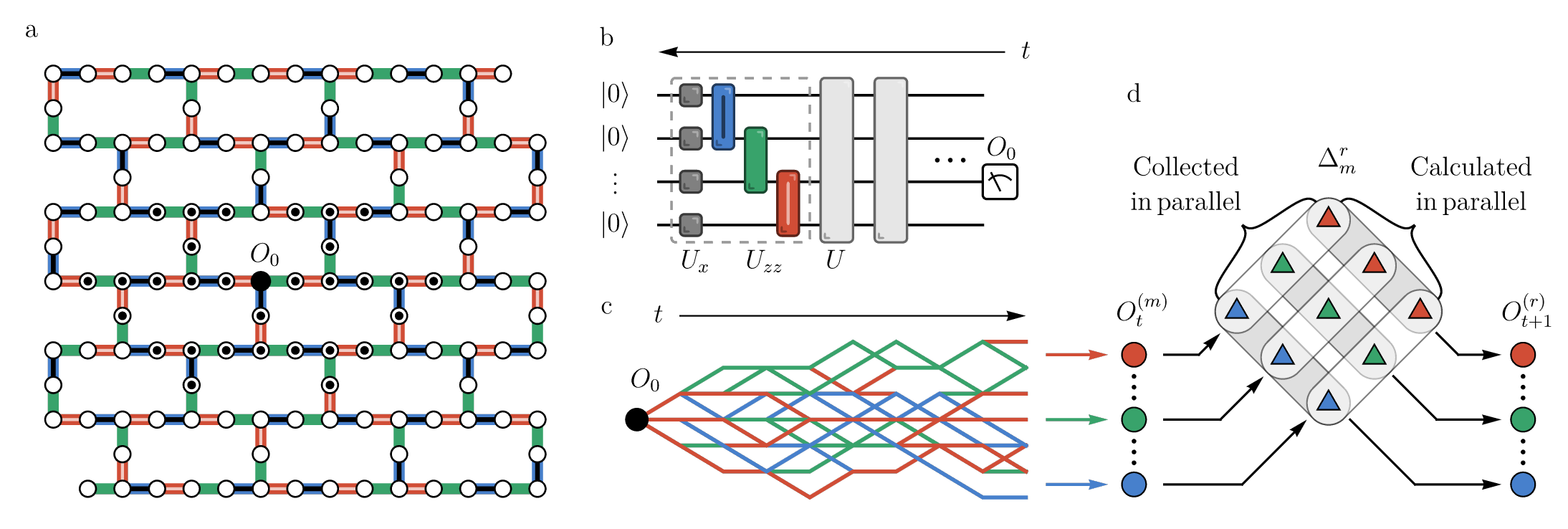}
    \caption{
        \textbf{Illustration of the model and simulation method.}
        (a) Heavy-hexagon geometry of the $127$-qubit IBM \textit{Eagle} processor simulated in this work. Circles represent qubits, with connections indicating where two-qubit gates can be applied. 
        The filled circle near the center marks the location of the initial operator $O_0=\sigma_z^{62}$, which is evolved in time. 
        Half-filled circles indicate the interior of the causal light cone after $t=5$ time steps, originating from the initial operator.
        The two-qubit connectivity is partitioned into three disjoint sets, each enabling parallel application of gates.
        (b) Quantum circuit implementing the Trotterized dynamics of the kicked Ising model. 
        Each layer $U$, enclosed by a dashed box, consists of a sequence of single-qubit gates $U_x$ and two-qubit entangling gates $U_{zz}$, forming a single time step. 
        The $U_{zz}$ gates are grouped into three color-coded sets, corresponding to the partitioning in (a), to facilitate parallel execution.  
        (c) Time evolution of the initial operator $O_0$, which begins as a single Pauli string and grows over time to include many additional Pauli strings, distributed across parallel processes labeled by $m$.
        In the Heisenberg picture, the operator evolution proceeds backward in time relative to the forward circuit direction shown in (b). 
        (d) At each time step $t$, the partial representations $O^{(m)}_t$ generate update lists $\Delta^r_m$, which are communicated in parallel to the target processes $r$. These updates are required to construct the next-step operator $O^{(r)}_{t+1}$.
        Note that the color coding in (a) and (b) is unrelated to that in (c) and (d).
    }  
    \label{parallel_figure}    
\end{figure*} 

In this work, we demonstrate a large-scale, high-performance implementation of the parallelized or-represented quantum algebra (ORQA) formalism~\cite{orqa}, executed on the supercomputer Fugaku.
ORQA enables efficient computation over large collections of Pauli strings and is naturally suited for quantum simulations involving operator dynamics, such as sparse Pauli evolution~\cite{begusic24,begusic24b}.
As a benchmark, we simulate the kicked Ising model on a two-dimensional heavy-hexagon geometry, following the layout used in the IBM \textit{Eagle} processor~\cite{kim23utility}. 
Our implementation tracks the dynamics of over one trillion Pauli strings, achieving a wall-clock time of only a few seconds per quantum gate.
This enables the simulation of circuits containing more than \num{10000} gates on $127$ qubits, with an overall runtime of just a few hours. 
Interestingly, we find that the relationship between simulation accuracy and the number of retained Pauli strings is highly sensitive to model-specific parameters. In particular, tuning a single physical parameter can significantly alter the complexity and effectiveness of the ORQA simulation. 

Our implementation demonstrates both strong and weak scaling up to $2^{17}$ parallel computational processes. 
Beyond this point, linearly increasing communication overhead becomes the dominant bottleneck and can only be mitigated when the number of Pauli strings exceeds the memory capacity of the available hardware.  
Within these practical limitations, our method enables the efficient simulation of arbitrary quantum circuits. 
As such, the ORQA framework constitutes a powerful and general-purpose numerical tool with broad applicability, particularly in quantum-hybrid high-performance computing and the simulation of quantum many-body dynamics. 

The remainder of this paper is organized as follows. In Section~\ref{sec:orqa}, we briefly review the ORQA formalism, introduce its key data structures, and summarize the parallel implementation. Section~\ref{sec:results} presents benchmark results for the kicked Ising model on the heavy-hexagon geometry and analyzes the performance characteristics of our method. We conclude in Section~\ref{sec:conclusion} with a summary of our findings and an outlook for future work. Additional details regarding the ORQA algorithm and parallel implementation are provided in Appendices~\ref{app:algorithm}--\ref{app:truncation}. Simulation accuracy is examined in Appendix~\ref{app:exact}, and the specifications of the computing system used for benchmarking are detailed in Appendix~\ref{app:device}.

\section{Methods} \label{sec:orqa}

The ORQA formalism~\cite{orqa} enables efficient manipulation of Pauli strings by reducing matrix multiplications to bitwise logical operations on encoded indices. 
A Pauli string on $n$ qubits is expressed as
\begin{equation} 
    \sigma_I = \sigma_{I_1|\dots|I_n} = \bigotimes_{i=1}^n \sigma_{I_i},
\end{equation}
where each $I_i$ is a two-bit string encoding the $i$th local Pauli operator according to the mapping $\sigma_{00_2}=\mathrm{id}$, $\sigma_{01_2}=\sigma_x$, $\sigma_{10_2}=\sigma_y$, and $\sigma_{11_2}=\sigma_z$. 
We refer to the full $2n$-bit concatenation $I=I_1|\dots|I_n$ as a multi-index.
The product of two arbitrary Pauli strings obeys the algebraic rule 
\begin{align}
    \sigma_I \sigma_J = i^{B(I,J)}\sigma_{I \veebar J},
    \label{orqaproduct}
\end{align}
where $\veebar$ denotes the bitwise exclusive-or (XOR) operation. 
The exponent $B(I,J)$ is determined by local contributions: 
\begin{equation}
    B(I,J) = \sum_{i=1}^n b(I_i,J_i),
\end{equation} 
where $I_i,J_i\in \{00_2,01_2,10_2,11_2\}$, and the local phase $b(I_i,J_i)$ captures the structure constants of the $\mathfrak{su}(2)$ algebra. 
Specifically, $b(I_i,J_i)=0$ whenever either $I_i=00_2$ or $J_i=00_2$ (i.e., an identity operator is present). Otherwise, the phase is given by $b(I_i,J_i)=\sum_{k=1}^3 \epsilon_{I_iJ_ik}$, where $\epsilon_{ijk}$ denotes the Levi-Cevita symbol.

To make use of this formalism, we express a generic Hermitian operator as a linear combination of Pauli strings:
\begin{equation}
    O=\sum_{I=0}^{4^n-1} O_I \sigma_I,
    \label{define_O}
\end{equation}
where $O_I\in\mathbb{R}$. 
In general, the summation runs over all $4^n$ possible Pauli strings on $n$ qubits.
However, in practice, a large fraction of the coefficients $O_I$ may vanish or be sufficiently small to be neglected without introducing significant error.
We denote by $|O|$ the number of Pauli strings retained in the representation of $O$, and refer to this quantity as the effective complexity, as it directly determines the computational cost of the algorithm. 

\begin{figure*}[h!t] 
    \includegraphics[width=1.0\linewidth]{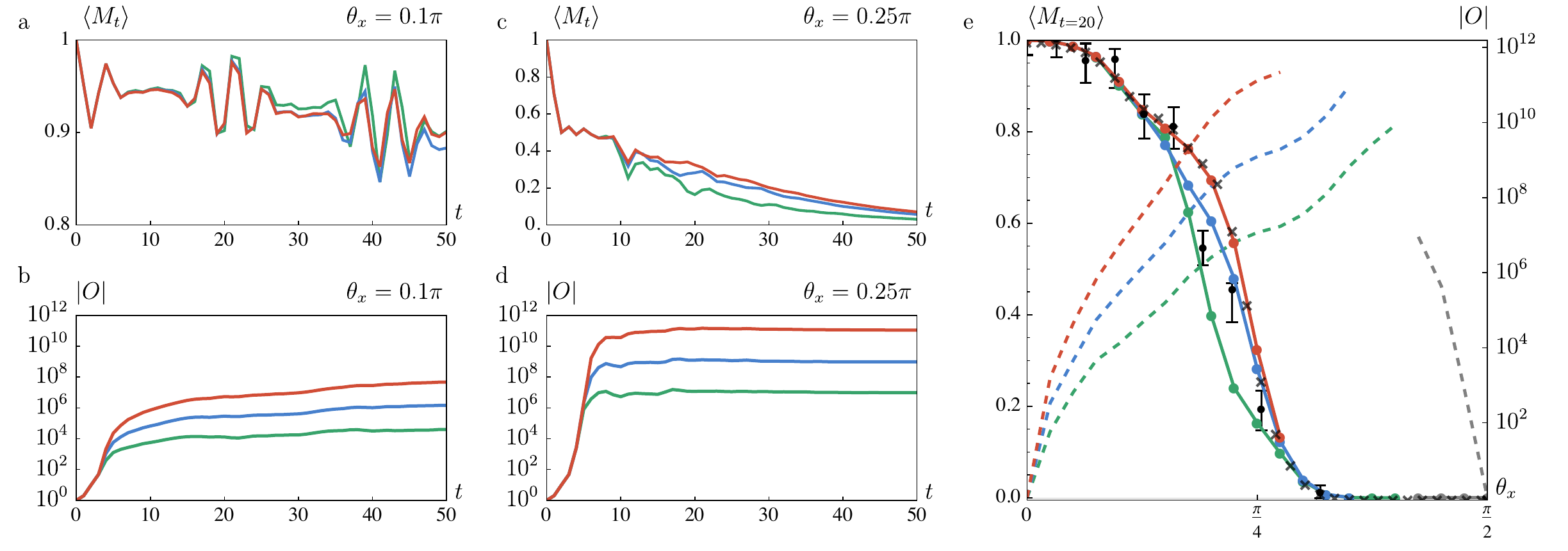} 
    \caption{  
        \textbf{Simulation of the kicked Ising model on the 127-qubit heavy-hexagon lattice.}
        (a) Time evolution of the magnetization $\langle M_t \rangle$ for $\theta_x=0.1\pi$ and truncation thresholds $\epsilon_0=10^{-4}$ (green), $10^{-5}$ (blue), and $10^{-6}$ (red). 
        (b) Corresponding effective complexity $|O|$ for the same simulations as in (a).
        (c,d) Same as (a) and (b), respectively, but for $\theta_x=0.25\pi$.
        (e) Magnetization $\langle M_t \rangle$ (solid circles) and effective complexity $|O|$ (dashed lines) at $t=20$ as functions of $\theta_x$ for different values of $\epsilon_0$: $10^{-2}$ (gray), 
        $10^{-4}$ (green), $10^{-5}$ (blue), and $10^{-6}$ (red). Error-mitigated experimental results from Ref.~\cite{kim23utility} are shown as black solid circles. For comparison, the results obtained using a tensor network method based on projected entangled pair operators, extrapolated to infinite bond dimension ($\chi\to\infty$), from Ref.~\cite{liao23} are shown as black crosses.  
        The underlying heavy-hexagon geometry and the location of the magnetization observable are shown in Fig.~\ref{parallel_figure}(a). 
    }  
    \label{lattices}    
\end{figure*} 

We consider simulations of quantum circuits in the Heisenberg picture, where the dynamics of observables $O$ measured at the end of the circuit are tracked. 
Using Eq.~(\ref{orqaproduct}), the unitary action of a quantum gate generated by a single operator $\sigma_J$ with rotation angle $\theta$, acting on $O$ as defined in Eq.~(\ref{define_O}), can be expressed as
\begin{widetext}
\begin{equation} 
    \frac{1}{2^{n}}\mathrm{Tr}\left[\sigma_I e^{i \frac{\theta}{2} \sigma_J } O e^{-i \frac{\theta}{2} \sigma_J }\right] = 
    \begin{cases}
        O_I, & \text{if } B(I,J) \text{ is even}, \\ 
        \cos(\theta)O_I + (-1)^{\frac{B(I,J)+1}{2}}\sin(\theta) O_{I\veebar J} , & \text{if } B(I,J) \text{ is odd},
    \end{cases}
    \label{update_rule}
\end{equation}
\end{widetext}
for all multi-indices $I$. 
Here, the left-hand side denotes the coefficient of $\sigma_I$ in the operator $O$ after conjugation by the unitary $e^{-i \frac{\theta}{2} \sigma_J }$, corresponding to the applied gate. Note that this expression relies on the identity $(I\veebar J)\veebar J=I$.
This update rule is analogous to those used in sparse Pauli dynamics~\cite{begusic24,begusic24b} and LOWESA~\cite{rudolph2023,fontana2023}, which evolve sums of Pauli strings using stabilizer decompositions~\cite{aaronson04} or Pauli transfer matrices~\cite{Chow2012, Rall2019}.
In contrast, our method is based on the ORQA formalism~\cite{orqa}, without requiring circuit preprocessing into Clifford and non-Clifford components.
Moreover, our implementation is optimized for execution on large-scale CPU-based supercomputers. Additional algorithmic details are provided in Appendix~\ref{app:algorithm}.

We parallelize this calculation by distributing the representation of the operator $O$ across $N$ parallel processes, each of which independently computes the updates given in Eq.~(\ref{update_rule}). 
The growth in the number of Pauli strings over time and their distribution across different processes are illustrated in Figs.~\ref{parallel_figure}(c) and \ref{parallel_figure}(d).
Details of the data structure used to represent $O$ numerically are provided in Appendix~\ref{app:hash}.
Each process, labeled by index $m$, maintains a partial representation $O^{(m)}$ of the full operator, such that 
\begin{equation}
    O=\sum_{m=0}^{N-1} O^{(m)}.
    \label{eq:Om}
\end{equation}

These partial representations are fully disjoint in the sense that each possible multi-index $I$ appears in exactly one of the $O^{(m)}$. 
To achieve and manage this efficiently, we introduce a distribution map
\begin{equation}
    f:\{0,1\}^{2n}\rightarrow \mathbb{N}_{N},
    \label{distributor}  
\end{equation}
which assigns each possible Pauli string to a specific process $m\in\mathbb{N}_{N}=\{0,1,\dots,N-1\}$. 
The exact form of $f$ is nontrivial and has a significant impact on parallel performance; our approach and its rationale are discussed in Appendix~\ref{app:distributionmap}. 
While the specific assignment of a given multi-index $I$ to a process $m$ is not directly important, 
it is essential that $f$ distributes multi-indices $I\in\{0,1\}^{2n}$ approximately uniformly across all processes.
This ensures balanced memory usage and computational load, i.e., $|O^{(m)}|\approx |O|N^{-1}$, which is crucial for maintaining scalability. 
Since each process holds a disjoint subset $O^{(m)}$, it must also compute a corresponding subset of updates, based on Eq.~(\ref{update_rule}), that are needed by other processes. 
Each process $m$ is therefore responsible for generating a collection of update lists, denoted $\Delta^r_m$, to be sent to processes $r\ne m$. 
Efficient computation and distribution of these updates across all processes is a central feature of our algorithm. 
Further implementation details are provided in Appendix~\ref{app:algorithm}. 

Within a circuit simulation, each quantum gate typically generates additional Pauli strings, leading to a steady--and potentially exponential--increase in the size of the operator representation $O$. 
To control this growth, we apply dynamic truncation to each partial representations $O^{(m)}$ by removing components $O^{(m)}_I$ that satisfy 
\begin{align}
   |O^{(m)}_I| &\leq \epsilon_0 \max_I |O_I|,  \label{truncation1}
\end{align}
where $\epsilon_0>0$ is a parameter that controls the strictness of the truncation.
Note that the maximum is taken over all coefficients in the full operator $O$, not just those in $O^{(m)}$, to ensure consistent truncation across all processes, regardless of how the distribution map $f$ partitions the data. 
In principle, various truncation strategies can be employed. 
We discuss the specific truncation scheme used in this work in Appendix~\ref{app:truncation}, and benchmark its accuracy through comparisons with exact numerical results in Appendix~\ref{app:exact}.

\section{Results} \label{sec:results}

To test our method, we consider the kicked Ising model, which has been studied experimentally~\cite{kim23utility} and whose dynamics have been semi-quantitatively reproduced using various classical simulation techniques~\cite{liao23,rudolph2023,Tindall24,Orus24,begusic23, begusic24}.
The unitary dynamics of this model are described by a Trotterized sequence of single- and two-qubit gates: 
\begin{align}
    U &= 
     \underbrace{\prod_{\langle i,j\rangle}\exp\left({-i\frac{\theta_{zz}}{2} \sigma_z^i\sigma_z^{j}}\right)}_{U_{zz}}\underbrace{\prod_{j=1}^n\exp\left({-i\frac{\theta_x}{2}  \sigma_x^j}\right)}_{U_x},
     \label{eq:u}
\end{align}
where $\sigma_x^j$ and $\sigma_z^j$ are the Pauli X and Pauli Z operators acting on the $j$th qubit. 
The two-qubit interactions are defined between neighboring qubits on a two-dimensional heavy-hexagon geometry with $n=127$ qubits, as illustrated in Fig.~\ref{parallel_figure}(a). The notation $\langle i,j\rangle$ in Eq.~(\ref{eq:u}) refers to a pair of adjacent qubits $i$ and $j$ on this lattice. 
The corresponding quantum circuit structure, consisting of layers of $U_{zz}$ and $U_x$ gates, is shown in Fig.~\ref{parallel_figure}(b). 
Throughout this work, we fix the interaction strength at $\theta_{zz}=-\frac{\pi}{2}$ and vary the transverse field parameter $\theta_x$.
Within this model, we study the local magnetization at qubit $j=62$ after applying $t$ layers of $U$, defined as 
\begin{equation} 
    \langle M_t \rangle = \langle 0 | (U^{\dagger})^t \sigma_z^{62} (U)^{t} | 0 \rangle,
    \label{magnetization}
\end{equation}
where $|0\rangle=| 0 \rangle^{\otimes n}$ is the initial product state.  
We compute this quantity in the Heisenberg picture by evolving the initial observable $O_0=\sigma_z^{62}$. 
We emphasize that, owing to the generality of our method, any linear combination of Pauli strings with arbitrary coefficients can be used as the initial operator $O_0$.
Likewise, the method is applicable to arbitrary quantum circuits $U$.
Details of the computing system used for our simulations are provided in Appendix~\ref{app:device}.

\begin{figure*}[ht!]
    \centering
    \includegraphics[width=0.9\linewidth]{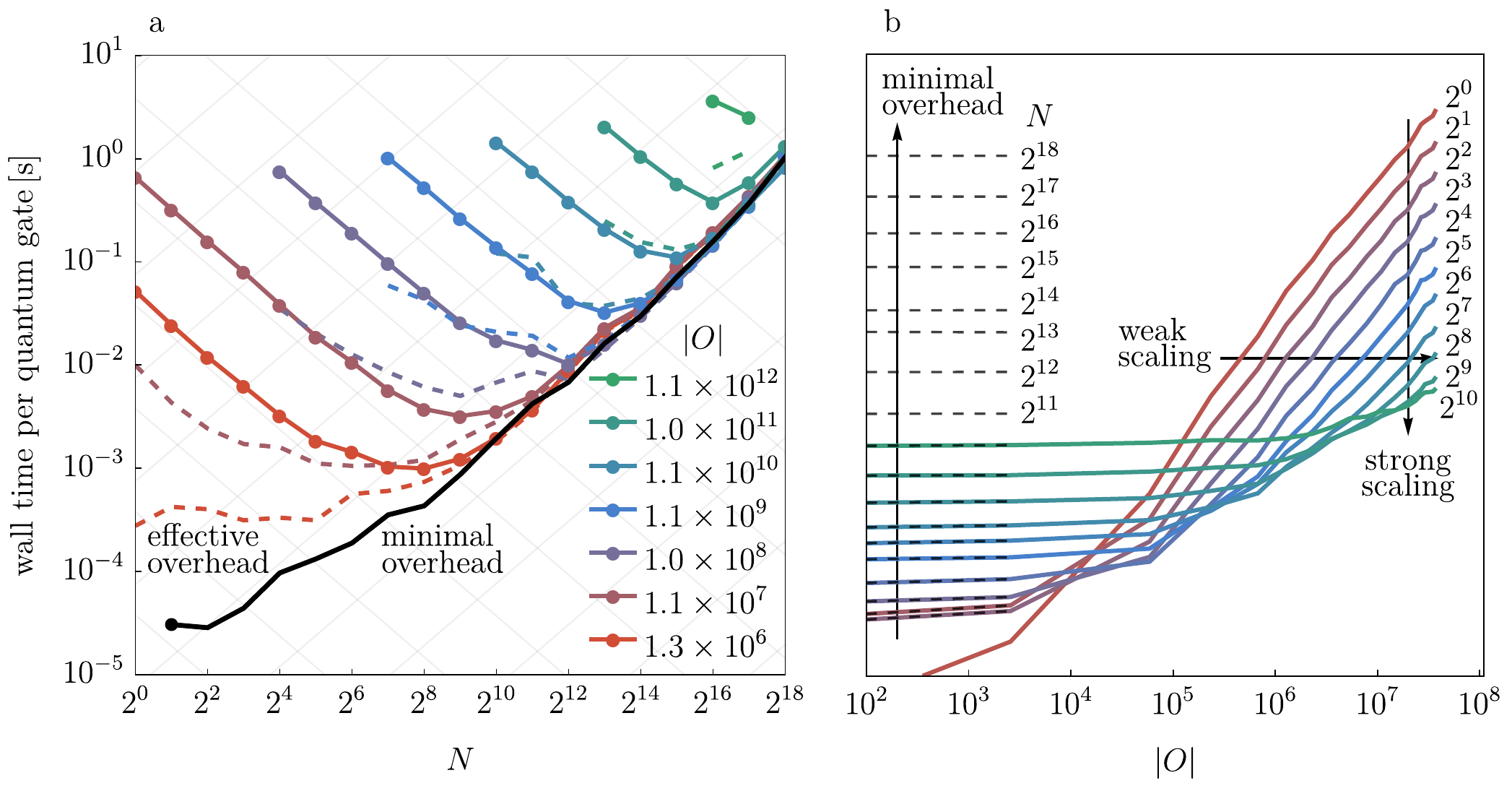}
    \caption{
    \textbf{Scaling behavior of quantum circuit simulations.}
    (a) Wall time per quantum gate as a function of the number of parallel computational processes $N$.  
    The solid black line denotes the minimal communication overhead that persists even in the limit $|O|\rightarrow 0$.
    Colored curves show the total wall time per gate (solid lines with symbols) and the effective communication overhead (dashed lines) for various effective complexities $|O|$, as labeled. 
    Diagonal grid lines represent ideal $N$ and $N^{-1}$ scaling for visual reference. 
    (b) Wall time per quantum gate as a function of effective complexity $|O|$ for different values of $N$, as indicated, using the same vertical-axis scale as in (a).
    For small $|O|$, performance is limited by the minimal communication overhead (dashed lines). As $|O|$ increases, the computational cost of simulating the operator dynamics becomes dominant, resulting in favorable scaling. 
    Strong scaling is evident from the approximate $N^{-1}$ scaling of wall time (vertical arrow), and the results also exhibit near-perfect weak scaling, with wall time remaining nearly constant when both $|O|$ and $N$ are scaled proportionally (horizontal arrow). 
    }
    \label{strong_scaling_figure}
\end{figure*}

We present the results of our simulations in Fig.~\ref{lattices} and Fig.~\ref{strong_scaling_figure}. 
Fig.~\ref{lattices}(a) shows the time evolution of the magnetization $\langle M_t \rangle$ for $\theta_x=0.1\pi$ and truncation thresholds $\epsilon_0=10^{-4}$ (green), $10^{-5}$ (blue), and $10^{-6}$ (red). 
The corresponding effective complexity $|O|$, defined as the number of retained Pauli strings, is shown in Fig.~\ref{lattices}(b) for the same set of parameters. 
Figs.~\ref{lattices}(c) and \ref{lattices}(d) present the same quantities for $\theta_x=0.25\pi$.
The magnetization results exhibit a visible dependence on the truncation threshold $\epsilon_0$, reflecting the impact of the number of Pauli strings retained at each stage of the evolution. 
As $\epsilon_0$ decreases, truncation becomes less aggressive, resulting in a systematic increase in $|O|$ and improved simulation accuracy. 
In all cases, the number of Pauli strings grows rapidly at early times, followed by a pronounced slowdown and eventual saturation due to truncation. 
This saturation behavior is qualitatively influenced by the choice of $\theta_x$, with different values leading to markedly different asymptotic values of $|O|$ for the same $\epsilon_0$. 
Nonetheless, even in regimes where saturation occurs, truncation-induced errors can accumulate over long time evolutions and eventually degrade simulation accuracy. 
This effect is particularly evident in Fig.~\ref{lattices}(a), where maintaining accuracy up to $t=50$ requires setting $\epsilon_0$ as low as $10^{-6}$.

Fig.~\ref{lattices}(e) shows the magnetization $\langle M_t \rangle$ and the effective complexity $|O|$ as functions of $\theta_x$ at a fixed evolution time $t=20$, for several values of the truncation threshold $\epsilon_0$. Error-mitigated experimental results from Ref.~\cite{kim23utility} are also shown as solid black circles.
Notably, the parameter regime $\theta_x\in[\pi/8, \pi/4]$ is particularly sensitive to truncation, exhibiting the largest deviations among the different values of  $\epsilon_0$. 
This challenging regime coincides with the region where other classical simulations also report substantial discrepancies, as discussed in recent studies (see, for example, Fig.~8 in Ref.~\cite{Tindall24}).
We additionally include the numerical results from Ref.~\cite{liao23}, shown as black crosses. 
These data were obtained using a projected entangled pair operator (PEPO) method, extrapolated to infinite bond dimension. The PEPO results are in excellent agreement with our simulation at $\epsilon_0=10^{-6}$, lending further support to the accuracy of our method in this regime.

For $\theta_x=0$ and $\theta_x=\pi/2$, the dynamics correspond to a fully classical regime and a dual-unitary circuit, respectively, both of which exhibit minimal effective complexity. 
As $\theta_x$ increases from zero, the effective complexity grows rapidly, reflecting the increasingly entangled and intricate nature of the system's dynamics. 
It is important to note, however, that a large number of Pauli strings does not necessarily indicate physically complex behavior. For example, when $\theta_x\lesssim \pi/2$, the system undergoes rapid thermalization, resulting in a fast decay of the magnetization. This qualitative feature is captured even under strong truncation, as shown for $\epsilon_0=10^{-2}$ in gray. In this regime, although the magnetization quickly decays and becomes featureless, the effective complexity still increases sharply as $\theta_x$ decreases from $\pi/2$, indicating that the growth in $|O|$ is not always directly linked to observable physical complexity. 

Fig.~\ref{strong_scaling_figure}(a) shows the wall time required to apply a single quantum gate as a function of the number of parallel computational processes $N$ during circuit simulation. This value is computed as the average wall time per circuit layer, obtained by dividing the total wall time for each layer by the the number of gates that it contains (127 single-qubit X-rotation gates and 144 two-qubit ZZ-rotation gates). 
The total wall times per gate are shown as solid lines with symbols.
Results are shown for various values of the effective complexity $|O|$ (i.e., the number of retained Pauli strings), corresponding to snapshots taken at different times during the time-evolution of the operator $O$ with varying parameters. 
The data exhibits clear strong-scaling behavior: the wall time approximately follows a $N^{-1}$ power-law scaling until communication overhead becomes the dominant cost. 
Dashed lines represent the measured communication overhead across processes. This includes both the actual data transfer time and the additional cost associated with initiating and finalizing MPI communication routines involving a large number of processes. 
We refer to this latter contribution as the minimal communication overhead, which is also indicated in isolation as a solid black line.
Importantly, when communication becomes the bottleneck, the actual data transfer time remains negligible, and nearly the entire overhead arises from this minimal initialization and finalization cost. 
As a result, we observe consistently favorable scaling across all tested values of $|O|$.
The minimal communication overhead is reported as an average, as it fluctuates slightly between runs depending on the physical allocation of processes and the topology of the interconnect network on the computing system.
At $N=2^{18}$, the minimal communication overhead becomes sufficiently large such that maintaining strong-scaling efficiency requires simulating more than $|O|=10^{12}$ Pauli strings. 
This value approaches the memory limits of our current hardware configuration, rendering further increases in $N$ ineffective under the present setup. 

Fig.~\ref{strong_scaling_figure}(b) shows the average wall time per quantum gate as a function of the effective complexity $|O|$. 
Each curve corresponds to the same circuit simulation executed with a different number of parallel computational processes $N$.
At early stages of the time evolution, the number of Pauli strings is relatively small, and the wall time is dominated by the minimal communication overhead, which is largely independent of $|O|$ but increases with $N$.
As the simulation progresses and the number of Pauli strings grows, the computational cost associated with updating the time-evolved operator $O$ becomes the dominant factor. 
In this regime, the wall time decreases approximately linearly with $N$, confirming strong-scaling behavior, as indicated by the vertical arrow.
Furthermore, the results demonstrate near-perfect weak scaling: the wall time remains nearly constant when both $|O|$ and $N$ are increased proportionally, as indicated by the horizontal arrow.

Finally, we emphasize that the scaling behavior shown in Fig.~\ref{strong_scaling_figure} is general and does not explicitly depend on physical details of the simulated system, such as the number of qubits.  
Instead, the overall performance is primarily dictated by the effective complexity $|O|$, which is governed by the chosen truncation strategy. 
By contrast, the relationship between $|O|$ and the accuracy of the simulation is determined by the physical characteristics of the system under study. Ultimately, it is these system-specific properties that determine the practical effectiveness of our method.

\section{Conclusion} \label{sec:conclusion}

We have presented a large-scale, parallelized, high-performance implementation of the ORQA formalism~\cite{orqa} for simulating unitary quantum dynamics in the Heisenberg picture, in the spirit of sparse Pauli dynamics~\cite{begusic23}. 
As a benchmark application, we simulated the kicked Ising model on a 127-qubit heavy-hexagon geometry, involving over \num{10000} quantum gates and up to $50$ circuit layers. 
Our implementation tracked over one trillion Pauli strings across $2^{17}$ parallel processes, achieving wall times of only a few seconds per gate, resulting in total runtimes of just a few hours on the supercomputer Fugaku. 
This performance highlights the scalability and numerical efficiency of our approach. 
Looking ahead, the ORQA framework is well suited for development on GPU-based supercomputing architectures, which promises to further accelerate performance. It also has the potential to play a vital role in quantum-HPC hybrid workflows on emerging quantum-centric supercomputing platforms. 

We have compared our simulation results for the magnetization $\langle M_t\rangle$ at $t=20$ across various values of $\theta_x$ with error-mitigated experimental data obtained from the IBM \textit{Eagle} processor~\cite{kim23utility}. Overall, the agreement is satisfactory, except in the region $\theta_x\approx 0.2\pi$--$0.25\pi$, which coincides with the regime where different classical simulation techniques yield divergent results, as comprehensively summarized in Ref.~\cite{Tindall24}. It is worth noting, however, that our results obtained with the smallest truncation threshold, $\epsilon_0=10^{-6}$, exhibit excellent agreement across all values of $\theta_x$ with those produced by a tensor network method based on PEPO in the Heisenberg picture, extrapolated to infinite bond dimension~\cite{liao23}. While we do not claim that our results are definitive, as systematic convergence checks with respect to $\epsilon_0$ remain to be performed, these comparisons suggest that further experimental investigations, particularly in this challenging parameter regime, would be highly valuable, in parallel with continued advancements in classical simulation techniques. 

Moreover, we have demonstrated both weak and strong scaling in our method, as well as a time complexity that scales nearly linearly with the number of retained Pauli strings. 
Importantly, this scaling behavior is independent of the specific physical system being simulated. 
The accuracy and tractability of simulations performed using our approach  depend solely on the number of Pauli strings that can be safely discarded without incurring significant error. 
This trade-off is governed by the physical properties of the system and the specific truncation strategy employed. 
We anticipate that future efforts aimed at tailoring truncation schemes to the structure of particular physical problems will play a central role in further improving both the efficiency and the scope of applicability of our algorithm.

The method presented in this work is general and highly versatile: it can, in principle, be applied directly to arbitrary quantum circuits, observables, and states, irrespective of the system's geometry or number of qubits. 
Although not explored in the present study, our implementation can be readily extended to simulate dissipative dynamics and imaginary-time evolution. 
By design, it also provides direct access to the expectation values of arbitrary Pauli strings without additional computational overhead, offering a notable advantage over many alternative approaches. 
Taken together, these features establish our implementation as a powerful and broadly applicable numerical tool for quantum many-body physics, with potential relevance to a wide range of problems in quantum simulation and quantum-HPC hybrid computation. 

\section*{Acknowledgments}

We are grateful to Hai-Jun Liao and Tao Xiang for sharing their results from Ref.~\cite{liao23}, and to the IBM Quantum team, as well as Tomonori Shirakawa, for valuable discussions. 
Numerical simulations were performed on the Supercomputer Fugaku and the HOKUSAI supercomputer at RIKEN (Project ID RB240003).
A portion of this work is based on results obtained from project JPNP20017, supported by the New Energy and Industrial Technology Development Organization (NEDO). 
This study was also supported by JSPS KAKENHI 
Grants 
No. JP21H04446 and
No. JP22K03520.
Additional support was provided by JST COI-NEXT (Grant No. JPMJPF2221) and the Program for Promoting Research on the Supercomputer Fugaku (Grant No. MXP1020230411) by MEXT, Japan.  
We further acknowledge support from the UTokyo Quantum Initiative, 
the RIKEN TRIP initiative (RIKEN Quantum), and
the COE research grant in computational science from Hyogo Prefecture and Kobe City through the Foundation for Computational Science.

\appendix 

\section{Algorithmic Details} 
\label{app:algorithm}

Our parallel numerical algorithm proceeds as follows. 
Each process $m$ holds a partial representation $O^{(m)}$ of the full operator $O$, such that their union reconstructs the complete operator, as described in Eq.~(\ref{eq:Om}).
Each $O^{(m)}$ is implemented as a dynamically sized list of tuples $(I,O^{(m)}_I)$, where $I$ is the multi-index labeling a Pauli string and $O^{(m)}_I$ is its corresponding coefficient. 
Each process sequentially applies the gates in the quantum circuit, where each gate is represented as a tuple $(J, \theta_J)$ corresponding to the unitary operator $e^{-i \frac{\theta_J}{2} \sigma_J }$.
For each gate, the process loops over the elements $(I, O^{(m)}_I)$ of its local operator $O^{(m)}$.
If $\sigma_I$ and $\sigma_J$ do not commute, the process registers the corresponding updates using Eq.~(\ref{update_rule}). 
Each process $m$ maintains $N$ dynamically allocated update lists, denoted $\Delta^{m}_{r}$, one for each process $r< N$.
The local update $O^{(m)}_I\leftarrow \cos(\theta_J)O^{(m)}_I$ is performed immediately. The associated non-local update is appended to the list $\Delta^{m}_{f(I\veebar J)}$ as a tuple 
\begin{equation}
(I\veebar J,\, (-1)^{\frac{B(I,J)+1}{2}}\sin(\theta_J) O^{(m)}_I), 
\end{equation}
where $f$ denotes the index-distribution function defined in Eq.~(\ref{distributor}). 
After all elements of $O^{(m)}$ have been processed, process $m$ will have generated update lists $\Delta^{m}_{r}$ for each $r$, representing the contributions that process $r$ must apply due to Pauli strings originally held by $m$. 
Each process then writes its update list $\Delta^{m}_{r}$ into a remote memory window associated with process $r$.
Finally, each process $m$ gathers and applies the updates addressed to it, i.e., the union $\cup_{r=0}^{N-1}\Delta^{r}_{m}$, to its local operator $O^{(m)}$. These procedures are schematically summarized in  Figs.~\ref{parallel_figure}(c) and \ref{parallel_figure}(d). 
This yields a globally consistent update to the operator $O$, with contributions uniformly distributed across all $N$ processes.

\begin{algorithm}[H]
\caption{Parallel ORQA Circuit Simulation}\label{alg:orqa}
\begin{algorithmic}
\Require Partial operator representation $O^{(m)}=\{(I,O^{(m)}_I), \dots\}$
\Require Quantum circuit $U=\{(J,\theta_J), \dots\}$
\Require Distribution map $f: \{0,1\}^{2n} \rightarrow \mathbb{N}_N$
\Require Truncation threshold $\epsilon$

\For{$(J, \theta_J)$ in $U$} \Comment{{Loop over quantum gates}}
    \State Initialize update buffers: $\Delta^m_r \gets \emptyset \quad \forall r \in  \mathbb{N}_N$
    \For{$(I, O^{(m)}_I)$ in $O^{(m)}$} \Comment{{Loop over Paulis trings}}
        \State $\phi \gets B(I, J)$
        \If{$\phi$ is odd} \Comment{{Calculate updates}}
            \State $O_I^{(m)} \leftarrow \cos(\theta_J) O_I^{(m)}$
            \State $\Delta^m_{f(I\veebar J)} \impliedby (I\veebar J, (-1)^{\frac{\phi+1}{2}} \sin(\theta_J) O_I^{(m)})$
        \EndIf
        \EndFor
        \State Allocate RMA windows: $\tilde{\Delta}^r_m \gets \emptyset \quad  \forall r \in  \mathbb{N}_N$ 
        \For{$r$ in $\mathbb{N}_N$} \Comment{{Distribute updates}}
            {\If{$\Delta^m_r \neq \emptyset$}
                \State {\textbf{MPI\_Put}} ${\Delta}^m_r \rightarrow \tilde{\Delta}^r_m$ 
            \EndIf}
        \EndFor
    \For{$(I, \delta)$ in $\cup_r \tilde{\Delta}^m_r$} \Comment{Apply received updates}
        \If{$I \notin O^{(m)}$}
            \State $O^{(m)} \impliedby (I, \delta_I)$
        \Else
            \State $O^{(m)}_I \leftarrow O^{(m)}_I + \delta_I$
        \EndIf
    \EndFor
    \For{$(I, O^{(m)}_I)$ in $O^{(m)}$} \Comment{{Truncate}}
        \If{$|O^{(m)}_I|\leq\epsilon$} 
                \State $O^{(m)} \implies (I, O^{(m)}_I)$ 
        \EndIf
    \EndFor
\EndFor
\end{algorithmic}
\end{algorithm}

In Algorithm~\ref{alg:orqa}, we present the pseudocode for our parallel method, describing the operations performed by process $m$ during a single layer of quantum gates. 
The algorithm captures the essential structure, including the sequence of computational and communication steps, while omitting certain implementation-specific details. In particular, we do not show the specifics of the truncation strategy, nor the procedure used to evaluate and broadcast the global maximum value of $|O_I|$ across all processes. 
We also omit technical aspects such as the additional communication required to allocate the appropriate buffer size for the remote memory access (RMA) window $\tilde\Delta$.
The symbols $\impliedby$ and $\implies$ are used to denote the insertion and deletion of elements in a list or hash map, respectively.

\section{Hash Map Representation} 
\label{app:hash}

An essential aspect of our implementation is the choice of data structure used to store the partial operator $O^{(m)}$. 
To this end, we employ a hash map. 
The suitability of a given hash map depends on several performance-critical criteria that vary by application. 
In our case, the data structure must satisfy the following three key requirements: First, it must support efficient read, write, and erase operations, even when managing more than $10^7$ elements. 
Second, it must incur minimal memory overhead, as memory usage constitutes a primary bottlenecks in large-scale simulations. 
Third, it must provide high iteration speed, since the algorithm loops over the entire contents of $O^{(m)}$ twice per quantum gate applied. 
While a custom-designed hash map tailored specifically to the ORQA formalism could in principle offer further performance gains, we found that the open-source library \textbf{emhash8}~\cite{emhash8} meets all three of the above requirements. Among the various alternatives that we tested, \textbf{emhash8} demonstrated the best overall performance in terms of memory efficiency, access speed, and iteration throughput. 

\section{Distribution Map}
\label{app:distributionmap}

The distribution map $f$ introduced in Eq.~(\ref{distributor}) deterministically assigns each Pauli string to one of the available computational processes, with the goal of achieving an approximately uniform distribution.  
A straightforward and viable choice for $f$ would be a deterministic, non-cryptographic hash function.
However, a more subtle yet equally important requirement is the sparsity of the update pattern $\Delta_r^m$--that is, each process $m$ should ideally communicate with only a small subset of other processes $r$, while still maintaining global load balance. 
This sparsity becomes increasingly crucial at large scales: as the number of processes grows, minimizing the total number of inter-process communication calls becomes more important than reducing the size of individual messages. Standard hash functions typically yield a uniform distribution of message sizes across all destinations, but they also tend to produce a large number of small messages sent to many different targets. This pattern can congest the communication network and introduce significant latency due to message handling overhead.
To mitigate this, we adopt a more structured, enumeration-based scheme for $f$ in place of a generic hash function. 
Details of our mapping strategy are described below. 

Given a multi-index $I$, we define the distribution map $f$ as  
\begin{equation}
    f(I) = \left[\sum_{j=0}^{{\lceil \frac{2n}{k} \rceil-1} } I^{(j,k)}\right]  \mod  N,
    \label{distributorf}
\end{equation}
where $n$ is the number of qubits, $N$ is the number of computational processes, and $k$ is an integer block size parameter chosen such that $2^k$ is ideally close to $N$.
Here, $I^{(j,k)}$ denotes the $j$th $k$-bit segment of the $2n$-bit multi-index $I$, counted from the least significant bits (i.e., from the right in our convention), and interpreted as an integer (e.g., $00_2=0$, $01_2=1$, $10_2=2$, and $11_2=3$ when $k=2$). 
This procedure partitions $I$ into $\lceil 2n/k \rceil$ blocks of $k$ bits, converts each block to an integer, sums the resulting values, and applies a modulo-$N$ reduction.  
This mapping approximately distributes all possible Pauli strings uniformly across the $N$ processes, while promoting communication sparsity--that is, each process tends to interact with only a small subset of other processes. 

To understand this, consider the set $P_m = \{I\in\{0,1\}^{2n} \mid f(I) = m \}$, which represents all possible Pauli strings assigned to process $m$.
When a quantum gate generated by $\sigma_J$ is applied, the multi-index is updated as $I \veebar J$, and the new target process becomes 
\begin{equation}
    f(I\veebar J) = m + \sum_{j=1}^{2|J|} \pm 2^{J_j \bmod k} \mod N,
\end{equation} 
where $|J|$ is the number of non-identity local Pauli operators in $\sigma_J$, and $J_j$ denotes the position of the $j$th such operator in the multi-index $J$.  

As an illustrative example, consider the 4-qubit Pauli string $\sigma_z\sigma_0\sigma_z\sigma_0$. Here, $|J|=2$ since two non-identity Pauli operators are present: $\sigma_z^{(1)}$ and $\sigma_z^{(3)}$, counting from the right. The full multi-index of this Pauli string is $11001100_2$, so that possible bit flips occur only at positions $J_1=2$, $J_2=3$, $J_3=6$, and $J_4=7$, again counting from the right and starting from $0$. 
In this case, the number of distinct target processes $f(I \veebar J)$ is bounded by $2^{2|J|}+1=17$ (depending on $N$ and $k$), which represents a substantial reduction compared to the $N$ possible destinations that would arise when using a typical hash function.
Thus, Eq.~\eqref{distributorf} achieves sparse inter-process communication while maintaining an approximately uniform global distribution of Pauli strings. 

\begin{figure}
    \centering
    \includegraphics[width=1\linewidth]{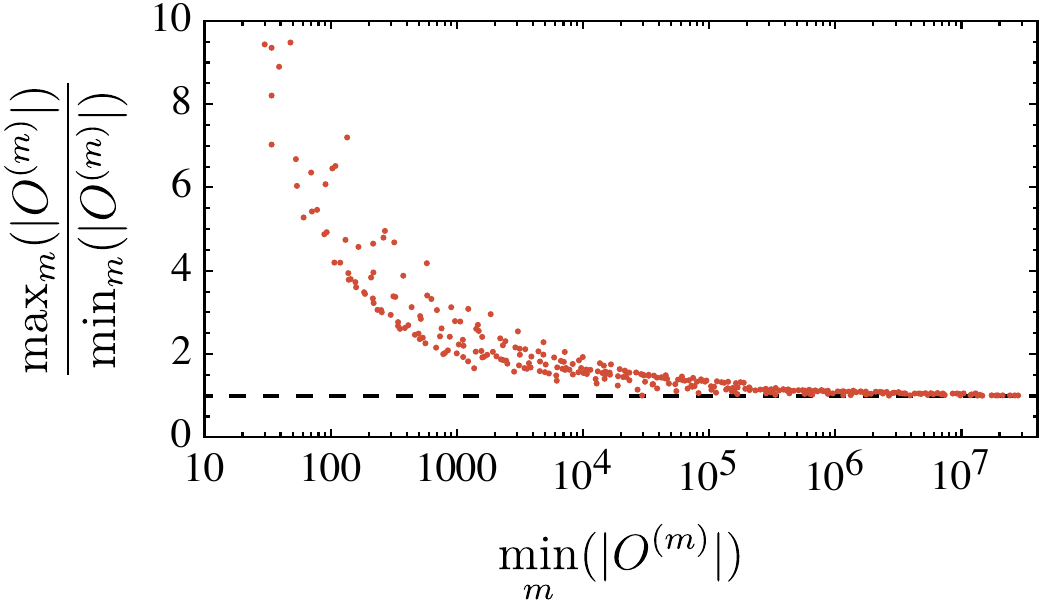}
    \caption{\textbf{Uniformity of Pauli string distribution across processes.}
    Ratio between the maximum and minimum number of Pauli strings among all partial representations $O^{(m)}$, plotted as a function of the minimum value $\min_m(|O^{(m)}|)$. 
    Values approaching unity indicate a more uniform distribution of Pauli strings across the parallel processes. 
    }
    \label{uniformity_figure}
\end{figure}

To demonstrate that the distribution map $f$ produces an overall uniform distribution of Pauli strings while maintaining communication sparsity, we present in Fig.~\ref{uniformity_figure} the ratio between the largest and smallest values of $|O^{(m)}|$ across all processes, plotted as a function of the minimum value $\min_m(|O^{(m)}|)$ observed in a representative simulation.
When this ratio approaches unity, the distribution of Pauli strings is considered effectively uniform. 
For small-scale simulations involving on the order of $10^3$ Pauli strings, this ratio can reach values as high as $4$, or even higher for fewer Pauli strings, indicating some degree of load imbalance. However, such imbalance is relevant only in the regime of very small problem sizes and has negligible impact on large-scale performance. 
As the number of Pauli strings per process increases, the ratio consistently converges toward $1$, indicating improved load balancing. In particular, once the number of retained Pauli strings exceeds approximately $10^5$ per process, the distribution becomes sufficiently uniform for practical purposes. 
In large-scale scenarios, such as simulations executed on the supercomputer Fugaku, it is advisable to target workloads involving between $10^6$ and $10^8$ Pauli strings per process. In this regime, the distribution achieved by the distribution map $f$ is effectively uniform, and the associated communication overhead becomes negligible. 

Although not employed in the present study, the uniformity of the distribution map can be further improved by modifying the baseline distribution map $f$ to a perturbed version: 
\begin{equation}
    g_s(I) = f(I) + \left( h(I)\bmod s\right) \mathrm{\,\,mod\,} N,
\end{equation}
where $h(I)$ is a deterministic, non-cryptographic hash function, and $s$ is a small positive integer. 
This extension introduces mild stochasticity into the output of $f$, effectively randomizing the distribution while largely preserving the communication sparsity provided by the original map. 
The limiting behavior of $g_s$ is well-controlled: when $s=1$, we recover the original mapping $g_1(I)=f(I)$; for $s=N$, the distribution approximates that of a conventional hash function. 
In this work, we adopt the unperturbed case $g_1(I) = f(I)$, which provides a favorable trade-off between communication sparsity and implementation simplicity. 
Nonetheless, choosing a small value of $s>1$ may offer improved load balancing in other settings, and exploring this direction could be valuable for future applications. 

\section{Truncation}
\label{app:truncation}

For a given physical problem, it is generally difficult to predict in advance how many Pauli strings must be retained in representing the time-evolved operator to achieve a desired level of accuracy. 
It is therefore essential to employ a truncation heuristic that selectively discards coefficients deemed negligible. 
Although such truncation inevitably introduces numerical error, it is crucial for keeping the effective complexity of the operator within computationally tractable bounds. 
The overarching objective is to maximize accuracy while minimizing the number of retained Pauli strings. 
We consider the development of more advanced truncation strategies to be a promising direction for further improving both the efficiency and accuracy of our method. 
As a baseline, in this work we analyze the simple amplitude-based criterion defined in Eq.~(\ref{truncation1}).

An interesting feature of the truncation procedure emerges in the regime where a large number of Pauli strings are retained: the distribution of their coefficients begins to exhibit stochastic characteristics. 
As the number of retained Pauli strings increases, the values of their coefficients tend to form a continuous distribution. 
To illustrate this behavior, we examine the normalized histogram of the coefficients $O_I$ in the time-evolved operator $O$, obtained from a representative simulation after $t$ circuit layers. We refer to this histogram as the Pauli string density, denoted $D_t(x)$, which characterizes the coefficient distribution in the presence of truncation.
To account for the overall decay in the magnitude of $O_I$ over time, we introduce a rescaled variable $x=O_I/\max_I|O_{I}|$. 
Under this normalization, the truncation threshold in Eq.~(\ref{truncation1}) corresponds to a sharp cutoff at $|x|=\epsilon_0$.

\begin{figure}[ht]
    \centering
    \includegraphics[width=1\linewidth]{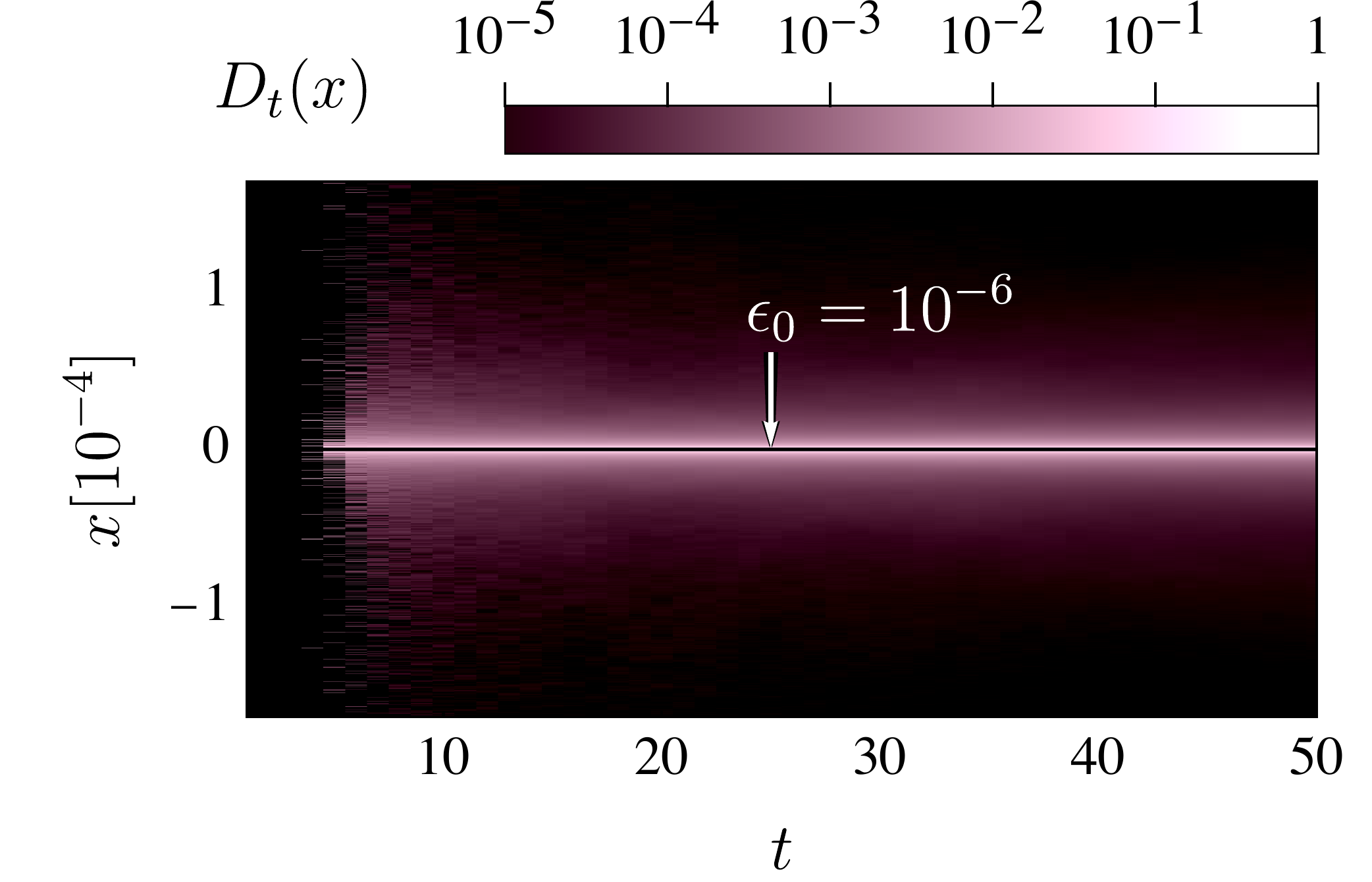}
    \caption{
    \textbf{Dynamics of the Pauli string density.}
    Normalized histogram $D_t(x)$ of the rescaled Pauli string coefficients $x=O_I/\max_I|O_I|$ as a function of time $t$, computed for $\theta_x=0.1\pi$ with truncation threshold $\epsilon_0=10^{-6}$. The sharp cutoff at $|x|\leq\epsilon_0$ reflects the applied truncation. }
    \label{histogram_figure}
\end{figure}

In Fig.~\ref{histogram_figure}, we present the Pauli string density $D_t(x)$ as a function of time $t$ for a representative quantum circuit simulation with $\theta_x=0.1\pi$ and a truncation threshold of $\epsilon_0=10^{-6}$.
At early times, the limited number of Pauli strings is evident from the appearance of discrete, high-density peaks in the histogram.
As the simulation progresses, the number of retained Pauli strings increases, and the histogram gradually evolves into a smoother, continuous distribution. 
A sharp cutoff appears at $|x|\leq \epsilon_0$, reflecting the imposed truncation threshold. 
The shape of the density distribution is governed by the dynamics of the underlying physical system and can vary significantly across different models. 
Consequently, the effectiveness of truncation is highly sensitive to the system-specific structure of the Pauli coefficient distribution. 
We anticipate that a deeper understanding of such distributions in various physical contexts will be essential for developing more effective and adaptive truncation strategies.

\section{Exact Solutions} \label{app:exact}

We validate the correctness of our algorithm by setting the truncation threshold to $\epsilon_0=0$ and comparing the results with those obtained from an exact state-vector simulation for a tractable system size.
Specifically, we simulate the first five time steps of the same system studied in Section~\ref{sec:results}, using $\theta_x=0.9\pi$. 
During this time evolution, the causal light cone reaches a maximum extent of 31 qubits, as illustrated in Fig.~\ref{parallel_figure}(a). 
Theoretically, our method becomes exact in the limit $\epsilon_0\to0$. 
Indeed, for $\epsilon_0=0$, we find that our results agree with the exact state-vector simulation within numerical double-precision accuracy (see Table~\ref{tab:my_label}).
Furthermore, even at a small but finite truncation threshold $\epsilon_0=10^{-5}$, the results remain in excellent agreement with the exact solution, as shown in Table~\ref{tab:my_label}. This confirms that the truncation procedure is correctly implemented and does not introduce spurious errors for sufficiently small values of $\epsilon_0$. 
\\

\begin{table}[H]
    \centering
    \begin{tabular}{c|c|c|c}
         $t$&state vector&$\epsilon_0=0$&$\epsilon_0=10^{-5}$ \\
         \hline
         $1$&-0.95105651629515&-0.95105651629515&-0.95105651629515 \\
         $2$&0.90450849718747&0.90450849718747&0.90450849718747 \\
         $3$&-0.9423841865631&-0.9423841865631&-0.9423841865631 \\
         $4$&0.97436799430392&0.97436799430396&0.97438562148434 \\
         $5$&-0.95315824248959&-0.95315824248965&-0.95311531321393 
    \end{tabular}
    \caption{Magnetization $\langle M_t \rangle$ computed using an exact state-vector simulation and our method with truncation thresholds $\epsilon_0=0$ and $\epsilon_0=10^{-5}$. Excellent agreement is observed for $\epsilon_0=0$, and the results for $\epsilon_0=10^{-5}$ remain close to the exact values, validating the accuracy of our implementation.}
    \label{tab:my_label}
\end{table}

\section{Computational Environment} \label{app:device}

{The simulations presented in this work were performed on the supercomputer Fugaku, which comprises 414 racks, each containing 384 compute nodes, for a total of \num{158976} nodes.
Each node is equipped with 48 computational cores and 32~GiB of memory.
Because memory usage constitutes a primary constraint in our implementation, we utilized between 8 and 22 computational processes per node, assigning one process per core.
This allocation provides each process with approximately 1--4~GiB of memory, sufficient to store on the order of $10^7$ Pauli strings, including all associated memory overhead. 
In our largest simulations, we employed up to \num{12288} nodes, corresponding to a total of up to $2^{18}=\num{262144}$ parallel processes.
This configuration allows us to efficiently simulate operator dynamics involving more than $10^{12}$ Pauli strings within practical wall-time limits. Remarkably, even at this scale, only a few percent of the total wall time is spent on inter-process communication, underscoring the scalability and efficiency of our implementation. 

\bibliography{lit.bib}

\begin{thebibliography}{31}%
\makeatletter
\providecommand \@ifxundefined [1]{%
 \@ifx{#1\undefined}
}%
\providecommand \@ifnum [1]{%
 \ifnum #1\expandafter \@firstoftwo
 \else \expandafter \@secondoftwo
 \fi
}%
\providecommand \@ifx [1]{%
 \ifx #1\expandafter \@firstoftwo
 \else \expandafter \@secondoftwo
 \fi
}%
\providecommand \natexlab [1]{#1}%
\providecommand \enquote  [1]{``#1''}%
\providecommand \bibnamefont  [1]{#1}%
\providecommand \bibfnamefont [1]{#1}%
\providecommand \citenamefont [1]{#1}%
\providecommand \href@noop [0]{\@secondoftwo}%
\providecommand \href [0]{\begingroup \@sanitize@url \@href}%
\providecommand \@href[1]{\@@startlink{#1}\@@href}%
\providecommand \@@href[1]{\endgroup#1\@@endlink}%
\providecommand \@sanitize@url [0]{\catcode `\\12\catcode `\$12\catcode `\&12\catcode `\#12\catcode `\^12\catcode `\_12\catcode `\%12\relax}%
\providecommand \@@startlink[1]{}%
\providecommand \@@endlink[0]{}%
\providecommand \url  [0]{\begingroup\@sanitize@url \@url }%
\providecommand \@url [1]{\endgroup\@href {#1}{\urlprefix }}%
\providecommand \urlprefix  [0]{URL }%
\providecommand \Eprint [0]{\href }%
\providecommand \doibase [0]{https://doi.org/}%
\providecommand \selectlanguage [0]{\@gobble}%
\providecommand \bibinfo  [0]{\@secondoftwo}%
\providecommand \bibfield  [0]{\@secondoftwo}%
\providecommand \translation [1]{[#1]}%
\providecommand \BibitemOpen [0]{}%
\providecommand \bibitemStop [0]{}%
\providecommand \bibitemNoStop [0]{.\EOS\space}%
\providecommand \EOS [0]{\spacefactor3000\relax}%
\providecommand \BibitemShut  [1]{\csname bibitem#1\endcsname}%
\let\auto@bib@innerbib\@empty
%</preamble>
\bibitem [{\citenamefont {Arute}\ \emph {et~al.}(2019)\citenamefont {Arute}, \citenamefont {Arya}, \citenamefont {Babbush}, \citenamefont {Bacon}, \citenamefont {Bardin}, \citenamefont {Barends}, \citenamefont {Biswas}, \citenamefont {Boixo}, \citenamefont {Brandao}, \citenamefont {Buell} \emph {et~al.}}]{arute2019quantum}%
  \BibitemOpen
  \bibfield  {author} {\bibinfo {author} {\bibfnamefont {F.}~\bibnamefont {Arute}}, \bibinfo {author} {\bibfnamefont {K.}~\bibnamefont {Arya}}, \bibinfo {author} {\bibfnamefont {R.}~\bibnamefont {Babbush}}, \bibinfo {author} {\bibfnamefont {D.}~\bibnamefont {Bacon}}, \bibinfo {author} {\bibfnamefont {J.~C.}\ \bibnamefont {Bardin}}, \bibinfo {author} {\bibfnamefont {R.}~\bibnamefont {Barends}}, \bibinfo {author} {\bibfnamefont {R.}~\bibnamefont {Biswas}}, \bibinfo {author} {\bibfnamefont {S.}~\bibnamefont {Boixo}}, \bibinfo {author} {\bibfnamefont {F.~G.}\ \bibnamefont {Brandao}}, \bibinfo {author} {\bibfnamefont {D.~A.}\ \bibnamefont {Buell}}, \emph {et~al.},\ }\bibfield  {title} {\bibinfo {title} {Quantum supremacy using a programmable superconducting processor},\ }\href {https://doi.org/10.1038/s41586-019-1666-5} {\bibfield  {journal} {\bibinfo  {journal} {Nature}\ }\textbf {\bibinfo {volume} {574}},\ \bibinfo {pages} {505} (\bibinfo {year} {2019})}\BibitemShut {NoStop}%
\bibitem [{\citenamefont {Liu}\ \emph {et~al.}(2021)\citenamefont {Liu}, \citenamefont {Liu}, \citenamefont {Li}, \citenamefont {Fu}, \citenamefont {Yang}, \citenamefont {Song}, \citenamefont {Zhao}, \citenamefont {Wang}, \citenamefont {Peng}, \citenamefont {Chen}, \citenamefont {Guo}, \citenamefont {Huang}, \citenamefont {Wu},\ and\ \citenamefont {Chen}}]{liu2021closing}%
  \BibitemOpen
  \bibfield  {author} {\bibinfo {author} {\bibfnamefont {Y.~A.}\ \bibnamefont {Liu}}, \bibinfo {author} {\bibfnamefont {X.~L.}\ \bibnamefont {Liu}}, \bibinfo {author} {\bibfnamefont {F.~N.}\ \bibnamefont {Li}}, \bibinfo {author} {\bibfnamefont {H.}~\bibnamefont {Fu}}, \bibinfo {author} {\bibfnamefont {Y.}~\bibnamefont {Yang}}, \bibinfo {author} {\bibfnamefont {J.}~\bibnamefont {Song}}, \bibinfo {author} {\bibfnamefont {P.}~\bibnamefont {Zhao}}, \bibinfo {author} {\bibfnamefont {Z.}~\bibnamefont {Wang}}, \bibinfo {author} {\bibfnamefont {D.}~\bibnamefont {Peng}}, \bibinfo {author} {\bibfnamefont {H.}~\bibnamefont {Chen}}, \bibinfo {author} {\bibfnamefont {C.}~\bibnamefont {Guo}}, \bibinfo {author} {\bibfnamefont {H.}~\bibnamefont {Huang}}, \bibinfo {author} {\bibfnamefont {W.}~\bibnamefont {Wu}},\ and\ \bibinfo {author} {\bibfnamefont {D.}~\bibnamefont {Chen}},\ }\bibfield  {title} {\bibinfo {title} {Closing the "quantum supremacy" gap: achieving real-time simulation of a random quantum circuit using a new sunway supercomputer},\ }in\ \href {https://doi.org/10.1145/3458817.3487399} {\emph {\bibinfo {booktitle} {Proceedings of the International Conference for High Performance Computing, Networking, Storage and Analysis}}},\ \bibinfo {series and number} {SC '21}\ (\bibinfo  {publisher} {Association for Computing Machinery},\ \bibinfo {address} {New York, NY, USA},\ \bibinfo {year} {2021})\BibitemShut {NoStop}%
\bibitem [{\citenamefont {Kim}\ \emph {et~al.}(2023)\citenamefont {Kim}, \citenamefont {Eddins}, \citenamefont {Anand}, \citenamefont {Wei}, \citenamefont {van~den Berg}, \citenamefont {Rosenblatt}, \citenamefont {Nayfeh}, \citenamefont {Wu}, \citenamefont {Zaletel}, \citenamefont {Temme},\ and\ \citenamefont {Kandala}}]{kim23utility}%
  \BibitemOpen
  \bibfield  {author} {\bibinfo {author} {\bibfnamefont {Y.}~\bibnamefont {Kim}}, \bibinfo {author} {\bibfnamefont {A.}~\bibnamefont {Eddins}}, \bibinfo {author} {\bibfnamefont {S.}~\bibnamefont {Anand}}, \bibinfo {author} {\bibfnamefont {K.~X.}\ \bibnamefont {Wei}}, \bibinfo {author} {\bibfnamefont {E.}~\bibnamefont {van~den Berg}}, \bibinfo {author} {\bibfnamefont {S.}~\bibnamefont {Rosenblatt}}, \bibinfo {author} {\bibfnamefont {H.}~\bibnamefont {Nayfeh}}, \bibinfo {author} {\bibfnamefont {Y.}~\bibnamefont {Wu}}, \bibinfo {author} {\bibfnamefont {M.}~\bibnamefont {Zaletel}}, \bibinfo {author} {\bibfnamefont {K.}~\bibnamefont {Temme}},\ and\ \bibinfo {author} {\bibfnamefont {A.}~\bibnamefont {Kandala}},\ }\bibfield  {title} {\bibinfo {title} {Evidence for the utility of quantum computing before fault tolerance},\ }\href {https://doi.org/10.1038/s41586-023-06096-3} {\bibfield  {journal} {\bibinfo  {journal} {Nature}\ }\textbf {\bibinfo {volume} {618}},\ \bibinfo {pages} {500} (\bibinfo {year} {2023})}\BibitemShut {NoStop}%
\bibitem [{\citenamefont {Acharya}\ \emph {et~al.}(2025)\citenamefont {Acharya} \emph {et~al.}}]{googleError}%
  \BibitemOpen
  \bibfield  {author} {\bibinfo {author} {\bibfnamefont {R.}~\bibnamefont {Acharya}} \emph {et~al.},\ }\bibfield  {title} {\bibinfo {title} {Quantum error correction below the surface code threshold},\ }\href {https://doi.org/10.1038/s41586-024-08449-y} {\bibfield  {journal} {\bibinfo  {journal} {Nature}\ }\textbf {\bibinfo {volume} {638}},\ \bibinfo {pages} {920} (\bibinfo {year} {2025})}\BibitemShut {NoStop}%
\bibitem [{\citenamefont {Preskill}(2018)}]{Preskill2018}%
  \BibitemOpen
  \bibfield  {author} {\bibinfo {author} {\bibfnamefont {J.}~\bibnamefont {Preskill}},\ }\bibfield  {title} {\bibinfo {title} {Quantum {C}omputing in the {NISQ} era and beyond},\ }\href {https://doi.org/10.22331/q-2018-08-06-79} {\bibfield  {journal} {\bibinfo  {journal} {{Quantum}}\ }\textbf {\bibinfo {volume} {2}},\ \bibinfo {pages} {79} (\bibinfo {year} {2018})}\BibitemShut {NoStop}%
\bibitem [{\citenamefont {Bharti}\ \emph {et~al.}(2022)\citenamefont {Bharti}, \citenamefont {Cervera-Lierta}, \citenamefont {Kyaw}, \citenamefont {Haug}, \citenamefont {Alperin-Lea}, \citenamefont {Anand}, \citenamefont {Degroote}, \citenamefont {Heimonen}, \citenamefont {Kottmann}, \citenamefont {Menke}, \citenamefont {Mok}, \citenamefont {Sim}, \citenamefont {Kwek},\ and\ \citenamefont {Aspuru-Guzik}}]{RevModPhysNisq}%
  \BibitemOpen
  \bibfield  {author} {\bibinfo {author} {\bibfnamefont {K.}~\bibnamefont {Bharti}}, \bibinfo {author} {\bibfnamefont {A.}~\bibnamefont {Cervera-Lierta}}, \bibinfo {author} {\bibfnamefont {T.~H.}\ \bibnamefont {Kyaw}}, \bibinfo {author} {\bibfnamefont {T.}~\bibnamefont {Haug}}, \bibinfo {author} {\bibfnamefont {S.}~\bibnamefont {Alperin-Lea}}, \bibinfo {author} {\bibfnamefont {A.}~\bibnamefont {Anand}}, \bibinfo {author} {\bibfnamefont {M.}~\bibnamefont {Degroote}}, \bibinfo {author} {\bibfnamefont {H.}~\bibnamefont {Heimonen}}, \bibinfo {author} {\bibfnamefont {J.~S.}\ \bibnamefont {Kottmann}}, \bibinfo {author} {\bibfnamefont {T.}~\bibnamefont {Menke}}, \bibinfo {author} {\bibfnamefont {W.-K.}\ \bibnamefont {Mok}}, \bibinfo {author} {\bibfnamefont {S.}~\bibnamefont {Sim}}, \bibinfo {author} {\bibfnamefont {L.-C.}\ \bibnamefont {Kwek}},\ and\ \bibinfo {author} {\bibfnamefont {A.}~\bibnamefont {Aspuru-Guzik}},\ }\bibfield  {title} {\bibinfo {title} {Noisy intermediate-scale quantum algorithms},\ }\href {https://doi.org/10.1103/RevModPhys.94.015004} {\bibfield  {journal} {\bibinfo  {journal} {Rev. Mod. Phys.}\ }\textbf {\bibinfo {volume} {94}},\ \bibinfo {pages} {015004} (\bibinfo {year} {2022})}\BibitemShut {NoStop}%
\bibitem [{\citenamefont {Alexeev}\ \emph {et~al.}(2024)\citenamefont {Alexeev}, \citenamefont {Amsler}, \citenamefont {Barroca}, \citenamefont {Bassini}, \citenamefont {Battelle}, \citenamefont {Camps}, \citenamefont {Casanova}, \citenamefont {Choi}, \citenamefont {Chong}, \citenamefont {Chung} \emph {et~al.}}]{Alexeev2024}%
  \BibitemOpen
  \bibfield  {author} {\bibinfo {author} {\bibfnamefont {Y.}~\bibnamefont {Alexeev}}, \bibinfo {author} {\bibfnamefont {M.}~\bibnamefont {Amsler}}, \bibinfo {author} {\bibfnamefont {M.~A.}\ \bibnamefont {Barroca}}, \bibinfo {author} {\bibfnamefont {S.}~\bibnamefont {Bassini}}, \bibinfo {author} {\bibfnamefont {T.}~\bibnamefont {Battelle}}, \bibinfo {author} {\bibfnamefont {D.}~\bibnamefont {Camps}}, \bibinfo {author} {\bibfnamefont {D.}~\bibnamefont {Casanova}}, \bibinfo {author} {\bibfnamefont {Y.~J.}\ \bibnamefont {Choi}}, \bibinfo {author} {\bibfnamefont {F.~T.}\ \bibnamefont {Chong}}, \bibinfo {author} {\bibfnamefont {C.}~\bibnamefont {Chung}}, \emph {et~al.},\ }\bibfield  {title} {\bibinfo {title} {Quantum-centric supercomputing for materials science: A perspective on challenges and future directions},\ }\href {https://doi.org/https://doi.org/10.1016/j.future.2024.04.060} {\bibfield  {journal} {\bibinfo  {journal} {Future Generation Computer Systems}\ }\textbf {\bibinfo {volume} {160}},\ \bibinfo {pages} {666} (\bibinfo {year} {2024})}\BibitemShut {NoStop}%
\bibitem [{\citenamefont {Robledo-Moreno}\ \emph {et~al.}(2024)\citenamefont {Robledo-Moreno}, \citenamefont {Motta}, \citenamefont {Haas}, \citenamefont {Javadi-Abhari}, \citenamefont {Jurcevic}, \citenamefont {Kirby}, \citenamefont {Martiel}, \citenamefont {Sharma}, \citenamefont {Sharma}, \citenamefont {Shirakawa} \emph {et~al.}}]{javier24}%
  \BibitemOpen
  \bibfield  {author} {\bibinfo {author} {\bibfnamefont {J.}~\bibnamefont {Robledo-Moreno}}, \bibinfo {author} {\bibfnamefont {M.}~\bibnamefont {Motta}}, \bibinfo {author} {\bibfnamefont {H.}~\bibnamefont {Haas}}, \bibinfo {author} {\bibfnamefont {A.}~\bibnamefont {Javadi-Abhari}}, \bibinfo {author} {\bibfnamefont {P.}~\bibnamefont {Jurcevic}}, \bibinfo {author} {\bibfnamefont {W.}~\bibnamefont {Kirby}}, \bibinfo {author} {\bibfnamefont {S.}~\bibnamefont {Martiel}}, \bibinfo {author} {\bibfnamefont {K.}~\bibnamefont {Sharma}}, \bibinfo {author} {\bibfnamefont {S.}~\bibnamefont {Sharma}}, \bibinfo {author} {\bibfnamefont {T.}~\bibnamefont {Shirakawa}}, \emph {et~al.},\ }\href {https://arxiv.org/abs/2405.05068} {\bibinfo {title} {Chemistry beyond exact solutions on a quantum-centric supercomputer}} (\bibinfo {year} {2024}),\ \Eprint {https://arxiv.org/abs/2405.05068} {arXiv:2405.05068 [quant-ph]} \BibitemShut {NoStop}%
\bibitem [{\citenamefont {Liao}\ \emph {et~al.}(2023)\citenamefont {Liao}, \citenamefont {Wang}, \citenamefont {Zhou}, \citenamefont {Zhang},\ and\ \citenamefont {Xiang}}]{liao23}%
  \BibitemOpen
  \bibfield  {author} {\bibinfo {author} {\bibfnamefont {H.-J.}\ \bibnamefont {Liao}}, \bibinfo {author} {\bibfnamefont {K.}~\bibnamefont {Wang}}, \bibinfo {author} {\bibfnamefont {Z.-S.}\ \bibnamefont {Zhou}}, \bibinfo {author} {\bibfnamefont {P.}~\bibnamefont {Zhang}},\ and\ \bibinfo {author} {\bibfnamefont {T.}~\bibnamefont {Xiang}},\ }\href {https://arxiv.org/abs/2308.03082} {\bibinfo {title} {Simulation of ibm's kicked ising experiment with projected entangled pair operator}} (\bibinfo {year} {2023}),\ \Eprint {https://arxiv.org/abs/2308.03082} {arXiv:2308.03082 [quant-ph]} \BibitemShut {NoStop}%
\bibitem [{\citenamefont {Rudolph}\ \emph {et~al.}(2023)\citenamefont {Rudolph}, \citenamefont {Fontana}, \citenamefont {Holmes},\ and\ \citenamefont {Cincio}}]{rudolph2023}%
  \BibitemOpen
  \bibfield  {author} {\bibinfo {author} {\bibfnamefont {M.~S.}\ \bibnamefont {Rudolph}}, \bibinfo {author} {\bibfnamefont {E.}~\bibnamefont {Fontana}}, \bibinfo {author} {\bibfnamefont {Z.}~\bibnamefont {Holmes}},\ and\ \bibinfo {author} {\bibfnamefont {L.}~\bibnamefont {Cincio}},\ }\href {https://arxiv.org/abs/2308.09109} {\bibinfo {title} {Classical surrogate simulation of quantum systems with lowesa}} (\bibinfo {year} {2023}),\ \Eprint {https://arxiv.org/abs/2308.09109} {arXiv:2308.09109 [quant-ph]} \BibitemShut {NoStop}%
\bibitem [{\citenamefont {Tindall}\ \emph {et~al.}(2024)\citenamefont {Tindall}, \citenamefont {Fishman}, \citenamefont {Stoudenmire},\ and\ \citenamefont {Sels}}]{Tindall24}%
  \BibitemOpen
  \bibfield  {author} {\bibinfo {author} {\bibfnamefont {J.}~\bibnamefont {Tindall}}, \bibinfo {author} {\bibfnamefont {M.}~\bibnamefont {Fishman}}, \bibinfo {author} {\bibfnamefont {E.~M.}\ \bibnamefont {Stoudenmire}},\ and\ \bibinfo {author} {\bibfnamefont {D.}~\bibnamefont {Sels}},\ }\bibfield  {title} {\bibinfo {title} {Efficient tensor network simulation of ibm's eagle kicked ising experiment},\ }\href {https://doi.org/10.1103/PRXQuantum.5.010308} {\bibfield  {journal} {\bibinfo  {journal} {PRX Quantum}\ }\textbf {\bibinfo {volume} {5}},\ \bibinfo {pages} {010308} (\bibinfo {year} {2024})}\BibitemShut {NoStop}%
\bibitem [{\citenamefont {Patra}\ \emph {et~al.}(2024)\citenamefont {Patra}, \citenamefont {Jahromi}, \citenamefont {Singh},\ and\ \citenamefont {Or\'us}}]{Orus24}%
  \BibitemOpen
  \bibfield  {author} {\bibinfo {author} {\bibfnamefont {S.}~\bibnamefont {Patra}}, \bibinfo {author} {\bibfnamefont {S.~S.}\ \bibnamefont {Jahromi}}, \bibinfo {author} {\bibfnamefont {S.}~\bibnamefont {Singh}},\ and\ \bibinfo {author} {\bibfnamefont {R.}~\bibnamefont {Or\'us}},\ }\bibfield  {title} {\bibinfo {title} {Efficient tensor network simulation of ibm's largest quantum processors},\ }\href {https://doi.org/10.1103/PhysRevResearch.6.013326} {\bibfield  {journal} {\bibinfo  {journal} {Phys. Rev. Res.}\ }\textbf {\bibinfo {volume} {6}},\ \bibinfo {pages} {013326} (\bibinfo {year} {2024})}\BibitemShut {NoStop}%
\bibitem [{\citenamefont {Begušić}\ and\ \citenamefont {Chan}(2023)}]{begusic23}%
  \BibitemOpen
  \bibfield  {author} {\bibinfo {author} {\bibfnamefont {T.}~\bibnamefont {Begušić}}\ and\ \bibinfo {author} {\bibfnamefont {G.~K.-L.}\ \bibnamefont {Chan}},\ }\href {https://arxiv.org/abs/2306.16372} {\bibinfo {title} {Fast classical simulation of evidence for the utility of quantum computing before fault tolerance}} (\bibinfo {year} {2023}),\ \Eprint {https://arxiv.org/abs/2306.16372} {arXiv:2306.16372 [quant-ph]} \BibitemShut {NoStop}%
\bibitem [{\citenamefont {Begušić}\ \emph {et~al.}(2024)\citenamefont {Begušić}, \citenamefont {Gray},\ and\ \citenamefont {Chan}}]{begusic24}%
  \BibitemOpen
  \bibfield  {author} {\bibinfo {author} {\bibfnamefont {T.}~\bibnamefont {Begušić}}, \bibinfo {author} {\bibfnamefont {J.}~\bibnamefont {Gray}},\ and\ \bibinfo {author} {\bibfnamefont {G.~K.-L.}\ \bibnamefont {Chan}},\ }\bibfield  {title} {\bibinfo {title} {Fast and converged classical simulations of evidence for the utility of quantum computing before fault tolerance},\ }\href {https://doi.org/10.1126/sciadv.adk4321} {\bibfield  {journal} {\bibinfo  {journal} {Science Advances}\ }\textbf {\bibinfo {volume} {10}},\ \bibinfo {pages} {eadk4321} (\bibinfo {year} {2024})},\ \Eprint {https://arxiv.org/abs/https://www.science.org/doi/pdf/10.1126/sciadv.adk4321} {https://www.science.org/doi/pdf/10.1126/sciadv.adk4321} \BibitemShut {NoStop}%
\bibitem [{\citenamefont {Biamonte}\ \emph {et~al.}(2017)\citenamefont {Biamonte}, \citenamefont {Wittek}, \citenamefont {Pancotti}, \citenamefont {Rebentrost}, \citenamefont {Wiebe},\ and\ \citenamefont {Lloyd}}]{QML17}%
  \BibitemOpen
  \bibfield  {author} {\bibinfo {author} {\bibfnamefont {J.}~\bibnamefont {Biamonte}}, \bibinfo {author} {\bibfnamefont {P.}~\bibnamefont {Wittek}}, \bibinfo {author} {\bibfnamefont {N.}~\bibnamefont {Pancotti}}, \bibinfo {author} {\bibfnamefont {P.}~\bibnamefont {Rebentrost}}, \bibinfo {author} {\bibfnamefont {N.}~\bibnamefont {Wiebe}},\ and\ \bibinfo {author} {\bibfnamefont {S.}~\bibnamefont {Lloyd}},\ }\bibfield  {title} {\bibinfo {title} {Quantum machine learning},\ }\href {https://doi.org/10.1038/nature23474} {\bibfield  {journal} {\bibinfo  {journal} {Nature}\ }\textbf {\bibinfo {volume} {549}},\ \bibinfo {pages} {195} (\bibinfo {year} {2017})}\BibitemShut {NoStop}%
\bibitem [{\citenamefont {Cerezo}\ \emph {et~al.}(2022)\citenamefont {Cerezo}, \citenamefont {Verdon}, \citenamefont {Huang}, \citenamefont {Cincio},\ and\ \citenamefont {Coles}}]{Cerezo22}%
  \BibitemOpen
  \bibfield  {author} {\bibinfo {author} {\bibfnamefont {M.}~\bibnamefont {Cerezo}}, \bibinfo {author} {\bibfnamefont {G.}~\bibnamefont {Verdon}}, \bibinfo {author} {\bibfnamefont {H.-Y.}\ \bibnamefont {Huang}}, \bibinfo {author} {\bibfnamefont {L.}~\bibnamefont {Cincio}},\ and\ \bibinfo {author} {\bibfnamefont {P.~J.}\ \bibnamefont {Coles}},\ }\bibfield  {title} {\bibinfo {title} {Challenges and opportunities in quantum machine learning},\ }\href {https://doi.org/10.1038/s43588-022-00311-3} {\bibfield  {journal} {\bibinfo  {journal} {Nature Computational Science}\ }\textbf {\bibinfo {volume} {2}},\ \bibinfo {pages} {567} (\bibinfo {year} {2022})}\BibitemShut {NoStop}%
\bibitem [{\citenamefont {White}(1992)}]{white1992density}%
  \BibitemOpen
  \bibfield  {author} {\bibinfo {author} {\bibfnamefont {S.~R.}\ \bibnamefont {White}},\ }\bibfield  {title} {\bibinfo {title} {Density matrix formulation for quantum renormalization groups},\ }\href {https://doi.org/10.1103/PhysRevLett.69.2863} {\bibfield  {journal} {\bibinfo  {journal} {Phys. Rev. Lett.}\ }\textbf {\bibinfo {volume} {69}},\ \bibinfo {pages} {2863} (\bibinfo {year} {1992})}\BibitemShut {NoStop}%
\bibitem [{\citenamefont {\"Ostlund}\ and\ \citenamefont {Rommer}(1995)}]{ostlund1995thermodynamic}%
  \BibitemOpen
  \bibfield  {author} {\bibinfo {author} {\bibfnamefont {S.}~\bibnamefont {\"Ostlund}}\ and\ \bibinfo {author} {\bibfnamefont {S.}~\bibnamefont {Rommer}},\ }\bibfield  {title} {\bibinfo {title} {Thermodynamic limit of density matrix renormalization},\ }\href {https://doi.org/10.1103/PhysRevLett.75.3537} {\bibfield  {journal} {\bibinfo  {journal} {Phys. Rev. Lett.}\ }\textbf {\bibinfo {volume} {75}},\ \bibinfo {pages} {3537} (\bibinfo {year} {1995})}\BibitemShut {NoStop}%
\bibitem [{\citenamefont {Schollwöck}(2011)}]{DMRG11}%
  \BibitemOpen
  \bibfield  {author} {\bibinfo {author} {\bibfnamefont {U.}~\bibnamefont {Schollwöck}},\ }\bibfield  {title} {\bibinfo {title} {The density-matrix renormalization group in the age of matrix product states},\ }\href {https://doi.org/https://doi.org/10.1016/j.aop.2010.09.012} {\bibfield  {journal} {\bibinfo  {journal} {Annals of Physics}\ }\textbf {\bibinfo {volume} {326}},\ \bibinfo {pages} {96} (\bibinfo {year} {2011})},\ \bibinfo {note} {january 2011 Special Issue}\BibitemShut {NoStop}%
\bibitem [{\citenamefont {Paeckel}\ \emph {et~al.}(2019)\citenamefont {Paeckel}, \citenamefont {Köhler}, \citenamefont {Swoboda}, \citenamefont {Manmana}, \citenamefont {Schollwöck},\ and\ \citenamefont {Hubig}}]{paeckel2019time}%
  \BibitemOpen
  \bibfield  {author} {\bibinfo {author} {\bibfnamefont {S.}~\bibnamefont {Paeckel}}, \bibinfo {author} {\bibfnamefont {T.}~\bibnamefont {Köhler}}, \bibinfo {author} {\bibfnamefont {A.}~\bibnamefont {Swoboda}}, \bibinfo {author} {\bibfnamefont {S.~R.}\ \bibnamefont {Manmana}}, \bibinfo {author} {\bibfnamefont {U.}~\bibnamefont {Schollwöck}},\ and\ \bibinfo {author} {\bibfnamefont {C.}~\bibnamefont {Hubig}},\ }\bibfield  {title} {\bibinfo {title} {Time-evolution methods for matrix-product states},\ }\href {https://doi.org/https://doi.org/10.1016/j.aop.2019.167998} {\bibfield  {journal} {\bibinfo  {journal} {Annals of Physics}\ }\textbf {\bibinfo {volume} {411}},\ \bibinfo {pages} {167998} (\bibinfo {year} {2019})}\BibitemShut {NoStop}%
\bibitem [{\citenamefont {Verstraete}\ and\ \citenamefont {Cirac}(2004{\natexlab{a}})}]{verstraete2004renormalization}%
  \BibitemOpen
  \bibfield  {author} {\bibinfo {author} {\bibfnamefont {F.}~\bibnamefont {Verstraete}}\ and\ \bibinfo {author} {\bibfnamefont {J.~I.}\ \bibnamefont {Cirac}},\ }\bibfield  {title} {\bibinfo {title} {Renormalization algorithms for quantum-many body systems in two and higher dimensions},\ }\href@noop {} {\bibfield  {journal} {\bibinfo  {journal} {arXiv preprint cond-mat/0407066}\ } (\bibinfo {year} {2004}{\natexlab{a}})}\BibitemShut {NoStop}%
\bibitem [{\citenamefont {Verstraete}\ and\ \citenamefont {Cirac}(2004{\natexlab{b}})}]{verstraete2004valence}%
  \BibitemOpen
  \bibfield  {author} {\bibinfo {author} {\bibfnamefont {F.}~\bibnamefont {Verstraete}}\ and\ \bibinfo {author} {\bibfnamefont {J.~I.}\ \bibnamefont {Cirac}},\ }\bibfield  {title} {\bibinfo {title} {Valence-bond states for quantum computation},\ }\href {https://doi.org/10.1103/PhysRevA.70.060302} {\bibfield  {journal} {\bibinfo  {journal} {Phys. Rev. A}\ }\textbf {\bibinfo {volume} {70}},\ \bibinfo {pages} {060302} (\bibinfo {year} {2004}{\natexlab{b}})}\BibitemShut {NoStop}%
\bibitem [{\citenamefont {Or{\'u}s}(2019)}]{Orus19}%
  \BibitemOpen
  \bibfield  {author} {\bibinfo {author} {\bibfnamefont {R.}~\bibnamefont {Or{\'u}s}},\ }\bibfield  {title} {\bibinfo {title} {Tensor networks for complex quantum systems},\ }\href {https://doi.org/10.1038/s42254-019-0086-7} {\bibfield  {journal} {\bibinfo  {journal} {Nature Reviews Physics}\ }\textbf {\bibinfo {volume} {1}},\ \bibinfo {pages} {538} (\bibinfo {year} {2019})}\BibitemShut {NoStop}%
\bibitem [{\citenamefont {Cirac}\ \emph {et~al.}(2021)\citenamefont {Cirac}, \citenamefont {P\'erez-Garc\'{\i}a}, \citenamefont {Schuch},\ and\ \citenamefont {Verstraete}}]{Cirac21}%
  \BibitemOpen
  \bibfield  {author} {\bibinfo {author} {\bibfnamefont {J.~I.}\ \bibnamefont {Cirac}}, \bibinfo {author} {\bibfnamefont {D.}~\bibnamefont {P\'erez-Garc\'{\i}a}}, \bibinfo {author} {\bibfnamefont {N.}~\bibnamefont {Schuch}},\ and\ \bibinfo {author} {\bibfnamefont {F.}~\bibnamefont {Verstraete}},\ }\bibfield  {title} {\bibinfo {title} {Matrix product states and projected entangled pair states: Concepts, symmetries, theorems},\ }\href {https://doi.org/10.1103/RevModPhys.93.045003} {\bibfield  {journal} {\bibinfo  {journal} {Rev. Mod. Phys.}\ }\textbf {\bibinfo {volume} {93}},\ \bibinfo {pages} {045003} (\bibinfo {year} {2021})}\BibitemShut {NoStop}%
\bibitem [{\citenamefont {Fontana}\ \emph {et~al.}(2023)\citenamefont {Fontana}, \citenamefont {Rudolph}, \citenamefont {Duncan}, \citenamefont {Rungger},\ and\ \citenamefont {Cîrstoiu}}]{fontana2023}%
  \BibitemOpen
  \bibfield  {author} {\bibinfo {author} {\bibfnamefont {E.}~\bibnamefont {Fontana}}, \bibinfo {author} {\bibfnamefont {M.~S.}\ \bibnamefont {Rudolph}}, \bibinfo {author} {\bibfnamefont {R.}~\bibnamefont {Duncan}}, \bibinfo {author} {\bibfnamefont {I.}~\bibnamefont {Rungger}},\ and\ \bibinfo {author} {\bibfnamefont {C.}~\bibnamefont {Cîrstoiu}},\ }\href {https://arxiv.org/abs/2306.05400} {\bibinfo {title} {Classical simulations of noisy variational quantum circuits}} (\bibinfo {year} {2023}),\ \Eprint {https://arxiv.org/abs/2306.05400} {arXiv:2306.05400 [quant-ph]} \BibitemShut {NoStop}%
\bibitem [{\citenamefont {Begušić}\ and\ \citenamefont {Chan}(2024)}]{begusic24b}%
  \BibitemOpen
  \bibfield  {author} {\bibinfo {author} {\bibfnamefont {T.}~\bibnamefont {Begušić}}\ and\ \bibinfo {author} {\bibfnamefont {G.~K.-L.}\ \bibnamefont {Chan}},\ }\href {https://arxiv.org/abs/2409.03097} {\bibinfo {title} {Real-time operator evolution in two and three dimensions via sparse pauli dynamics}} (\bibinfo {year} {2024}),\ \Eprint {https://arxiv.org/abs/2409.03097} {arXiv:2409.03097 [quant-ph]} \BibitemShut {NoStop}%
\bibitem [{\citenamefont {Broers}\ and\ \citenamefont {Mathey}(2024)}]{orqa}%
  \BibitemOpen
  \bibfield  {author} {\bibinfo {author} {\bibfnamefont {L.}~\bibnamefont {Broers}}\ and\ \bibinfo {author} {\bibfnamefont {L.}~\bibnamefont {Mathey}},\ }\href {https://arxiv.org/abs/2404.09312} {\bibinfo {title} {Exclusive-or encoded algebraic structure for efficient quantum dynamics}} (\bibinfo {year} {2024}),\ \Eprint {https://arxiv.org/abs/2404.09312} {arXiv:2404.09312 [cond-mat.other]} \BibitemShut {NoStop}%
\bibitem [{\citenamefont {Aaronson}\ and\ \citenamefont {Gottesman}(2004)}]{aaronson04}%
  \BibitemOpen
  \bibfield  {author} {\bibinfo {author} {\bibfnamefont {S.}~\bibnamefont {Aaronson}}\ and\ \bibinfo {author} {\bibfnamefont {D.}~\bibnamefont {Gottesman}},\ }\bibfield  {title} {\bibinfo {title} {Improved simulation of stabilizer circuits},\ }\href {https://doi.org/10.1103/PhysRevA.70.052328} {\bibfield  {journal} {\bibinfo  {journal} {Phys. Rev. A}\ }\textbf {\bibinfo {volume} {70}},\ \bibinfo {pages} {052328} (\bibinfo {year} {2004})}\BibitemShut {NoStop}%
\bibitem [{\citenamefont {Chow}\ \emph {et~al.}(2012)\citenamefont {Chow}, \citenamefont {Gambetta}, \citenamefont {C\'orcoles}, \citenamefont {Merkel}, \citenamefont {Smolin}, \citenamefont {Rigetti}, \citenamefont {Poletto}, \citenamefont {Keefe}, \citenamefont {Rothwell}, \citenamefont {Rozen}, \citenamefont {Ketchen},\ and\ \citenamefont {Steffen}}]{Chow2012}%
  \BibitemOpen
  \bibfield  {author} {\bibinfo {author} {\bibfnamefont {J.~M.}\ \bibnamefont {Chow}}, \bibinfo {author} {\bibfnamefont {J.~M.}\ \bibnamefont {Gambetta}}, \bibinfo {author} {\bibfnamefont {A.~D.}\ \bibnamefont {C\'orcoles}}, \bibinfo {author} {\bibfnamefont {S.~T.}\ \bibnamefont {Merkel}}, \bibinfo {author} {\bibfnamefont {J.~A.}\ \bibnamefont {Smolin}}, \bibinfo {author} {\bibfnamefont {C.}~\bibnamefont {Rigetti}}, \bibinfo {author} {\bibfnamefont {S.}~\bibnamefont {Poletto}}, \bibinfo {author} {\bibfnamefont {G.~A.}\ \bibnamefont {Keefe}}, \bibinfo {author} {\bibfnamefont {M.~B.}\ \bibnamefont {Rothwell}}, \bibinfo {author} {\bibfnamefont {J.~R.}\ \bibnamefont {Rozen}}, \bibinfo {author} {\bibfnamefont {M.~B.}\ \bibnamefont {Ketchen}},\ and\ \bibinfo {author} {\bibfnamefont {M.}~\bibnamefont {Steffen}},\ }\bibfield  {title} {\bibinfo {title} {Universal quantum gate set approaching fault-tolerant thresholds with superconducting qubits},\ }\href {https://doi.org/10.1103/PhysRevLett.109.060501} {\bibfield  {journal} {\bibinfo  {journal} {Phys. Rev. Lett.}\ }\textbf {\bibinfo {volume} {109}},\ \bibinfo {pages} {060501} (\bibinfo {year} {2012})}\BibitemShut {NoStop}%
\bibitem [{\citenamefont {Rall}\ \emph {et~al.}(2019)\citenamefont {Rall}, \citenamefont {Liang}, \citenamefont {Cook},\ and\ \citenamefont {Kretschmer}}]{Rall2019}%
  \BibitemOpen
  \bibfield  {author} {\bibinfo {author} {\bibfnamefont {P.}~\bibnamefont {Rall}}, \bibinfo {author} {\bibfnamefont {D.}~\bibnamefont {Liang}}, \bibinfo {author} {\bibfnamefont {J.}~\bibnamefont {Cook}},\ and\ \bibinfo {author} {\bibfnamefont {W.}~\bibnamefont {Kretschmer}},\ }\bibfield  {title} {\bibinfo {title} {Simulation of qubit quantum circuits via pauli propagation},\ }\href {https://doi.org/10.1103/PhysRevA.99.062337} {\bibfield  {journal} {\bibinfo  {journal} {Phys. Rev. A}\ }\textbf {\bibinfo {volume} {99}},\ \bibinfo {pages} {062337} (\bibinfo {year} {2019})}\BibitemShut {NoStop}%
\bibitem [{\citenamefont {ktprime}\ \emph {et~al.}(2025)\citenamefont {ktprime} \emph {et~al.}}]{emhash8}%
  \BibitemOpen
  \bibfield  {author} {\bibinfo {author} {\bibnamefont {ktprime}} \emph {et~al.},\ }\href@noop {} {\bibinfo {title} {emhash - fast and memory efficient open addressing c++ flat hash table/map}},\ \bibinfo {howpublished} {\url{https://github.com/ktprime/emhash}} (\bibinfo {year} {2025})\BibitemShut {NoStop}%
\end{thebibliography}%

\end{document}